\documentclass[prl,twocolumn,aps,floats]{revtex4}
\usepackage{graphicx}
\usepackage{psfrag}

\begin{document}

\title{Reply to Ryff's Comment on ``Experimental Nonlocality Proof of
Quantum Teleportation and Entanglement Swapping''}

\author{Thomas Jennewein$^1$, Gregor Weihs$^{1,2}$, Jian-Wei Pan$^1$, and Anton Zeilinger$^1$}
\affiliation{$^1$Institut f\"ur Experimentalphysik, Universit\"at Wien, Boltzmanngasse 5, A--1090 Wien, Austria\\
$^2$Ginzton Laboratory, S-23 Stanford University, Stanford, CA
94304, USA}


\begin{abstract}

Ryff's Comment raises the question of the meaning of the quantum
state. We argue that the quantum state is just the representative
of information available to a given observer. Then Ryff's
interpretation of one of our experiments and our original one are
both admissible.

\end{abstract}


\maketitle

Ryff's criticism \cite{1} of our interpretation of part of our
recent experiment \cite{2} on the teleportation of entanglement
gives us the opportunity to present our position in more detail.
The basic issue is the role of relative time order of various
detection events on the one hand and the meaning of a quantum
state on the other hand.

In the experiment two pairs of entangled photons are produced and
one photon from each pair is sent to Alice. The other photon from
each pair is sent to Bob (this might actually be two - even
spacelike -separated places). Bob is free to choose which
polarizations to  measure on his two photons separately. Likewise
Alice is free to choose whether she wants to project her two
photons onto an entangled state and thus effect quantum
teleportation or measure them individually. Most importantly,
each one of them decides which measurement to perform and
registers the results without being aware at all what kind of
measurement the other performs at which time.  Both Alice's data
and Bob's data are completely independent of whatever the other
decides to measure.

Then they ask themselves how their data are to be interpreted.
Obviously both Alice's and Bob's interpretations depend
critically on the information they have. It is assumed they both
know the initial entangled states. Alice then, on the basis of
her measurement result can make certain statements about Bob's
possible results. These can be collected into expectation
catalogs that give lists of results Bob may obtain for the
specific observables he might choose to measure. The quantum
state is no more than a most compact representative of such
expectation catalogs \cite{3}. If Alice decides to perform a
Bell-state analysis she will use an entangled state for her
prediction of Bob's results. If she measures the polarizations of
the two photons separately, she will use an unentangled product
state. In both cases she will be able to arrive at a correct
(maximal and in general probabilistic) set of predictions in both
being compatible with Bob's results. In the first case she
concludes, certainly correctly, that Bob's two photons are
entangled and teleportation has succeeded. In the second case she
will conclude that there is no entanglement between Bob's two
photons and that no teleportation has happened. But, as stressed
above, the data obtained by Bob are independent of Alice's
actions. Indeed his data set taken alone is completely random.

Likewise Bob will always use a product state based on his
measurement results and he thus will be able to predict Alice's
results both for the case when she performs a Bell state
measurement and when she does not.

It is now important to analyze what we mean by "prediction". As
the relative time ordering of Alice's and Bob's events is
irrelevant, "prediction" cannot refer to the time order of the
measurements. It is helpful to remember that the quantum state is
just an expectation catalog. Its purpose is to make predictions
about possible measurement results a specific observer does not
know yet. Thus which state is to be used depends on which
information Alice and Bob have and "prediction" means prediction
about measurement results they will learn in the future
independent of whether these measurements have already been
performed by someone or not. Also, in our point of view it is
irrelevant whether Alice performs her measurement earlier in any
reference frame than Bob's or later or even if they are spacelike
separated when the seemingly paradoxical situation arises that
different observers are  spacelike separated. In all these cases
Alice will use the same quantum state to predict the results she
will learn from Bob.

In short, we don't see any problem with Alice using her results
to predict which kind of results she will learn from Bob even if
he might already have obtained these results. There is no action
into the past since the events observed by Bob are independent of
which measurements Alice performs and at which time.

Thus we have no disagreement with Ryff's way to interpret our
experiment. But we certainly disagree with his position that his
way of looking at the situation is the only possible one. Yet we
would still agree with Peres \cite{4} that there is a possible
paradox here. But this paradox does not arise if the quantum
state is viewed to be no more than just a representative of
information.

\end{document}